\documentstyle [12pt]{article}
\newcommand{\cs}[3]{{{#3} \brace {#1 #2}}}

\newcommand{\h}[1]{\mathop{\lambda}\limits_{#1}\ \!\!\!}

\newcommand{\edf}{\ {\mathop{=}\limits^{\rm def}}\ }

\begin{document}
\begin{center}
{\bf {Path Deviation Equations in AP-Geometry}}
 \end{center}
\begin{center}
\bf{M.I.Wanas\footnote{Astronomy Department, Faculty of Science,
Cairo University, Giza,
 Egypt\\E-mail:wanas@frcu.eun.eg} \&
 M.E. Kahil\footnote{Mathematics Department,The American University in Cairo, Cairo, Egypt\\ E-mail:kahil@aucegypt.edu}}
\end{center}
 \abstract{Recently, it has been shown that Absolute Parallelism (AP)
     geometry admits paths that are naturally quantized. These paths
     have been used to describe the motion of spinning particles
     in a background gravitational field. In case of a weak static
     gravitational field limits, the paths are applied successfully to
     interpret the discrepancy in the motion of thermal neutrons
     in the Earth's gravitational field (COW-experiment). The aim
     of the present work is to explore the properties of the deviation
     equations corresponding to  these paths. In the present work
     the deviation equations are derived and compared to the
     geodesic deviation equation of the Riemannian geometry.}

\section{Introduction}
 The equation of path deviation plays an important role in geometry and its applications.
  In Riemannian geometry the equation of geodesic deviation is used to study, covariantly,
  the problem of stability [1], the problem of fixing the curvature constant of the universe [2] and other problems. Bazanski [3] has suggested a scheme, using which one can derive both the geodesic and geodesic deviation equations from one Lagrangian which can be written in the form
$$
L \edf g_{\mu \nu} U^{\mu} \frac{D \Psi^{\nu} }{DS}       \eqno{(1.1)}
$$
where $g_{\mu \nu}$ is the metric tensor, $U^{\mu}$ is a unit
tangent to the geodesic, $\Psi^{\nu}$ is the deviation vector and
$S$ is an affine parameter, The relation between the prefix
operator $\frac{D}{DS}$ and the covariant infix operator (;) can
be written, as usual, as
$$
 \frac{D \Psi^{\nu} }{DS}= \Psi^{\nu}_{~;\mu}U^{\mu}. \eqno{(1.2)}
$$
In Riemannian geometry, there is only one affine connection, Christoffel symbol. Thus the covariant derivative $\Psi^{\nu}_{~;\mu}$ is uniquely defined  and one can easily deduce that the definition (1.1) is unique.

In Absolute Parallelism (AP) geometry (cf. [4]) there are at least
three affine connections. These are the non-symmetric connection
$\Gamma^{\alpha}_{.\mu \nu}$, its dual
$\tilde{\Gamma}^{\alpha}_{.\mu \nu} ( = \Gamma^{\alpha}_{.\nu
\mu}) $ and its symmetric part $\Gamma^{\alpha}_{.( \mu \nu )}$.
In this geometry, a second order symmetric tensor $( g_{\mu \nu}
)$, which can play the role of the metric tensor, could be
defined. Consequently, Christoffel symbols could be defined as a
result of a metricity condition. Using these four affine
connections , the following tensor derivatives could be defined:
$$
 A^{\mu}_{\  ; \nu} \edf A^{\mu}_{~, \nu} + A^{\alpha} \cs{\alpha}{\nu}{\mu} ,    \eqno{(1.3)}
 $$
$$
 A^{\mu}_{+|~ \nu} \edf A^{\mu}_{~, \nu} +
A^{\alpha}~{\Gamma}^{\mu}_{.~\alpha \nu},\eqno{(1.4)}
$$

$$
 A^{\mu}_{.~ |~ \nu} \edf
A^{\mu}_{~, \nu} + A^{\alpha}~{\Gamma}^{\mu} _{.~(\alpha \nu)},\eqno{(1.5)}
$$
$$
 A^{\mu}_{-| \ \nu} \edf A^{\mu}_{\ , \nu} + A^{\alpha}
{\tilde{\Gamma}^{\mu}_{.~ \alpha \nu}}, \eqno{(1.6)}
$$
where $A^{\mu}$ is an arbitrary contravariant vector. The Bazanski scheme has been generalized [5] in AP-geometry, by using the above tensor derivatives in place of the covariant derivative used in the Lagrangian (1.1). Varying the generalized Lagrangian w.r.t. the deviation vector , a new set of path equations has been obtained. The general form of this set can be written as
$$
\frac{d^{2}x^{\mu}}{d^{2} \tau}+~a~\cs{\alpha}{\beta}{\mu} \frac{dx^{\alpha}}{d \tau} \frac{dx^{\beta}}{d\tau}~=-b~\Lambda_{(\alpha \beta ).}^{~~~~\mu}\frac{dx^{\alpha}}{d \tau} \frac{dx^{\beta}}{d\tau}   \eqno{(1.7)}
$$
where $\Lambda^{\alpha}_{. \mu \nu}$ is the torsion of the AP-space, $\tau$ is a parameter varying along the path, and $a, b$ are numerical parameters whose values are given in the following table. \\
\begin{center}
  {Table 1. Numerical Parameters of the New Set of Path Equations }      \end{center}
\begin{center}
\begin{tabular}{|c|c|c|} \hline
 ~~~~~~~~Connection used~~~~~~~~&~~~~~~~a~~~~~~~&~~~~~~~b~~~~~~~  \\
 \hline
& & \\
  ${\Gamma}^{\mu}_{.~\alpha \nu} $ & 1&  1      \\
\hline
& & \\
   ${\Gamma}^{\mu}_{.~ (\alpha \nu )}$ &1 &  $\frac{1}{2}$   \\
   \hline
& & \\
  $\tilde{\Gamma}^{\mu}_{.~\alpha\nu } $ & 1&  0       \\
 \hline
\end{tabular}
\end{center}
\vspace{0.5cm}
 From this table it is clear that the parameter of the torsion term in (1.7) has some quantum features [6].

It is the aim of the present work to explore the consequences of varying the same Lagrangian functions , used in the previous work [5], [6] w.r.t.
the tangent vector. This is done in order to study the impact of the quantum properties , appeared in the path equations, on the deviation equations
\section{Path Deviation Equations}
In this section, we are going to use the generalized Lagrangian
functions, used before [5], to derive the corresponding path
deviation equations. These functions can be written in the form:
$$ L \edf g_{\alpha \beta}~ U^{\alpha}~ {\frac{{{D}}{\Psi}^{\beta}}{{ {D}}
S}}~~~~~,  \eqno{(2.1)} $$
 $$ L^{0} \edf g_{\alpha \beta}~ W^{\alpha}~ {\frac{{{D}}{\zeta}^{\beta}}{{
{D}} S^{0}}}~~~~~,  \eqno{(2.2)} $$
$$ L^{+} \edf g_{\alpha \beta}~V^{\alpha}~ {\frac{{{D}}{\xi}^{\beta}}{{ {D}}
S^{+}}}~~~~~,  \eqno{(2.3)} $$
 $$ L^{-} \edf g_{\alpha \beta}~ J^{\alpha}~ {\frac{{{D}}{\eta}^{\beta}}{{
{D}} S^{-}}}~~~~~,  \eqno{(2.4)} $$
where
${\xi}^{\beta}, {\zeta}^{\beta}$ and ${\eta}^{\beta}$ are the
vectors giving the deviation from the curves characterized by the
evolution parameters $S^{+}, S^{0}$ and $S^{-}$ respectively.
Varying (2.1) w.r.t. the unit vector $U^{\mu}$, we obtain , after applying an action principle,
$$
\frac{D^{2}{\Psi}^{\alpha}}{D S^{2}} + R^{~~~\alpha}_{\beta \mu  . \rho}~ U^{\mu}\Psi^{\beta}U^{\rho} =~ 0, \eqno{(2.5)}
$$
which is the geodesic deviation equation of the Riemannian space associated with the AP-space. The tensor $R^{\alpha}_{.\mu\nu\sigma}$ is the Riemann-Christoffel curvature tensor defined using Christoffel symbols. Varying the Lagrangian (2.2) w.r.t the vector $W^{\sigma}$ and using an action principle we get,
$$
\frac{D^{2}\zeta^{\alpha}}{D S^{o~2}} + N^{~~\alpha}_{\beta \mu. \rho}~ W^{\mu}\zeta^{\beta}W^{\rho} =~ \frac{1}{2} \Lambda_{\nu \rho .}^{~~~\alpha}\frac{D \zeta^{\nu}}{D S^{o}}W^{\rho}~+~ \gamma^{\alpha}_{. \rho \nu}\frac{D \zeta^{\nu}}{D S^{o}} W^{\rho} , \eqno{(2.6)}
$$
where $N^{\alpha}_{.\mu \nu \sigma}$ is a curvature tensor
defined, using the symmetric part of the AP-connection, as
$$
N^{\alpha}_{. \mu \nu \sigma} \edf \Gamma^{\alpha}_{( \mu \sigma), \nu} -
\Gamma^{\alpha}_{( \mu \nu ),\sigma} + \Gamma^{\epsilon}_{( \mu \sigma)} \Gamma^{\alpha}_{( \epsilon \nu)} - \Gamma^{\epsilon}_{( \mu \nu)} \Gamma^{\alpha}_{( \epsilon \sigma)} . \eqno{(2.7)}$$
Again, varying the Lagrangian (2.3) w.r.t. the vector $V^{\sigma}$ and using an action principle we get,
$$
\frac{D^{2}\xi^{\alpha}}{D S^{+~2}}  =  \Lambda_{\nu \rho
.}^{~~~\alpha}\frac{D\xi^{\nu}}{D S^{+}} V^{\rho}~ . \eqno{(2.8)}
$$
The absence of the curvature term from this equation is due to the
identical vanishing of the curvature tensor, corresponding to the
connection $\Gamma^{\alpha}_{\mu  \nu}$ (cf. [4]),
$$
M^{\alpha}_{. \mu \nu \sigma} \edf \Gamma^{\alpha}_{ \mu \sigma, \nu} -
\Gamma^{\alpha}_{ \mu \nu ,\sigma} + \Gamma^{\epsilon}_{ \mu \sigma} \Gamma^{\alpha}_{ \epsilon \nu} - \Gamma^{\epsilon}_{ \mu \nu} \Gamma^{\alpha}_{ \epsilon \sigma} \equiv 0 . \eqno{(2.9)}
$$
Finally, varying the Lagrangian(2.4) w.r.t. $J^{\sigma}$ we get,
after using an action principle, the equation
$$
\frac{D^{2}\eta^{\alpha}}{D S^{-~2}} + \tilde{M}^{~~~\alpha}_{ \beta \mu . \rho}~J^{\mu}\eta^{\beta}J^{\rho}   =  2 \gamma^{\alpha}_{. \rho \nu}\frac{D \eta^{\nu}}{D S^{-}} J^{\rho}, \eqno{(2.10)}
$$
where $\tilde{M}^{\alpha}_{.\mu \nu \sigma}$ is a curvature tensor defined using the dual affine connection $\tilde{\Gamma}^{\alpha}_{ \mu \nu}$,
$$
\tilde{M}^{\alpha}_{. \mu \nu \sigma} \edf \tilde{\Gamma}^{\alpha}_{. \mu \sigma, \nu} -
\tilde{\Gamma}^{\alpha}_{. \mu \nu ,\sigma} + \tilde{\Gamma}^{\epsilon}_{. \mu \sigma} \tilde{\Gamma}^{\alpha}_{. \epsilon \nu} - \tilde{\Gamma}^{\epsilon}_{. \mu \nu} \tilde{\Gamma}^{\alpha}_{. \epsilon \sigma} .  \eqno{(2.11)}
$$
\section{Relation Between The Curvature Tensors}
 In order to compare the path deviation equations obtained, in section 2, it is convenient to relate the curvature tensors appeared in the equations. The non-symmetric connection ${\Gamma}^{\alpha}_{ \mu \nu}$ is related to the Christoffel symbols by the relation
(cf.[4])
$$
\Gamma^{\alpha}_{. \mu \nu} = ~\cs{\mu}{\nu}{\alpha}+~\gamma^{\alpha}_{. \mu \nu} ,
 \eqno{(3.1)}
$$
where,
$$
\gamma^{\alpha}_{. \mu \nu} \edf \h{i}^{\alpha} \h{i}_{\mu ;\nu}  , \eqno{(3.2)}
$$
and $\h{i}_{\mu}$ are the
tetrad vectors giving the structure of the AP-space. Substituting from (3.1) into definition (2.7) we get,
$$
N^{\lambda}_{. \mu \nu \sigma} = \frac{1}{2}( R^{\lambda}_{. \mu \nu \sigma}+ \tilde{Q}^{\lambda}_{. \mu \nu \sigma} + \Lambda^{\epsilon}_{. \mu  \sigma } \Lambda^{\lambda}_{. \nu  \epsilon}) , \eqno{(3.3)}
$$
where,
$$
\tilde{Q}^{\lambda}_{. \mu \nu \sigma} \edf \gamma^{\lambda}_{. \sigma \mu ;\nu} -\gamma^{\lambda}_{. \nu \mu ; \sigma}+ \gamma^{\epsilon}_{. \sigma \mu} \gamma^{\lambda}_{ \nu \epsilon} - \gamma^{\epsilon}_{. \nu \mu} \gamma^{\lambda}_{.\sigma \epsilon}  . \eqno{(3.4)}
$$
As it is well known that $R^{\lambda}_{. \mu \nu \sigma}$ is skew symmetric in the last two indices $\nu$ \& $\sigma$, and from definitions (2.7), (3.4) it is clear that $N^{\lambda}_{. \mu \nu \sigma}$ and $\tilde{Q}^{\lambda}_{. \mu \nu \sigma}$ are skew symmetric in last two indices $\nu$ \& $\sigma$. Consequently, the symmetric part of the last term in (3.3) vanishes identically, and the tensor can written as
$$
N^{\lambda}_{. \mu \nu \sigma} = \frac{1}{2}(R^{\lambda}_{. \mu \nu \sigma}+ \tilde{Q}^{\lambda}_{. \mu \nu \sigma} + \Lambda^{\epsilon}_{. \mu [ \sigma } \Lambda^{\lambda}_{. \nu ]  \epsilon} ) . \eqno{(3.5)}
$$
Again substituting from (3.1) into definition (2.11) we get after
some reductions,
$$
\tilde{M}^{\alpha}_{. \mu \nu \sigma} \edf R^{\alpha}_{. \mu \nu \sigma}+ \tilde{Q}^{\alpha}_{. \mu \nu \sigma} . \eqno{(3.6)}
$$
 Now the path deviation equations (2.8), (2.6) and (2.10) can be written, respectively,
 as

$$
\frac{D^{2}\xi^{\alpha}}{D S^{+~2}}   =  \Lambda_{\nu \rho
.}^{~~~\alpha}\frac{D\xi^{\nu}}{D S^{+}} V^{\rho}~ , \eqno{(3.7)}
$$

$$
\frac{D^{2}\zeta^{\alpha}}{D S^{o~2}} +\frac{1}{2} ( R^{~~~\alpha}_{\beta \mu . \rho}~+~ \tilde{Q}^{~~~\alpha}_{\beta \mu . \rho} +g^{\alpha \sigma} \Lambda^{\epsilon}_{. \mu [ \rho } \Lambda_{\beta \sigma ]  \epsilon} ) W^{\mu}\zeta^{\beta}W^{\rho} =~ \frac{1}{2} \Lambda_{\nu \rho .}^{~~~\alpha}\frac{D \zeta^{\nu}}{D S^{o}} W^{\rho}~+~ \gamma^{\alpha}_{. \rho \nu}\frac{D \zeta^{\nu}}{D S^{o}} W^{\rho} , \eqno{(3.8)}
$$
$$
\frac{D^{2}\eta^{\alpha}}{D S^{-~2}} + ( R^{~~~\alpha}_{\beta \mu . \rho}~+~ \tilde{Q}^{~~~\alpha}_{\beta \mu . \rho}) ~J^{\mu}\eta^{\beta}J^{\rho}   =  2 \gamma^{\alpha}_{. \rho \nu}\frac{D \eta^{\nu}}{D S^{-}} J^{\rho} . \eqno{(3.9)}
$$
This set of equations can be written into the general form
$$
\frac{D^{2}\Psi^{\alpha}}{D \tau^{2}} +  a ( R^{~~~\alpha}_{\beta \mu . \rho}~+~ \tilde{Q}^{~~~\alpha}_{\beta \mu . \rho}+ 2(1-a)g^{\alpha \sigma} \Lambda^{\epsilon}_{. \mu [ \rho } \Lambda_{\beta \sigma ]  \epsilon} ) U^{\mu}\Psi^{\beta}U^{\rho} =~ b \Lambda_{\nu \rho .}^{~~~\alpha}\frac{D \Psi^{\nu}}{D \tau} U^{\rho}~+~c \gamma^{\alpha}_{. \rho \nu}\frac{D \Psi^{\nu}}{D \tau} U^{\rho} . \eqno{(3.10)}
$$
\section{Discussion and Concluding Remarks}
In the present work, we have derived a new set of path deviation equations in AP-geometry. This new set can be written in the general form (3.10), from which it is clear that this form contains four terms with different numerical parameters.
 These terms are the second derivative of the deviation vector, the curvature term,
  the torsion term and the contortion term. Comparing these equations with the geodesic deviation equation (2.5),
   we construct the following table which gives the values of the parameters $a$,$b$ \& $c$.
   \\ \\ \\
\begin{center}
{Table 2. Values of the Parameters of (3.10) } \\
\end{center}
\begin{center}
\begin{tabular}{|c|c|c|c|} \hline
 ~~~~Connection used~~~~  &~~~~ a~~~~ &~~~~ b~~~~ &~~~~ c~~~~  \\
 \hline
& & &\\
 ${\Gamma}^{\mu}_{.~\alpha \nu} $ &~0~&~1~&~0    \\
\hline
& & &\\
  ${\Gamma}^{\mu}_{.~ (\alpha \nu )}$ &~$\frac{1}{2}~$ &~  $\frac{1}{2}$ ~& ~1      \\
   \hline
& & & \\
  $\tilde{\Gamma}^{\mu}_{.~\alpha\nu } $ &~1~&~0~&~2      \\
   \hline
\end{tabular}
\end{center}
\vspace{0.5cm} From the treatment given in the present work, and
the above table,
we note that:\\
1- The disappearance of the torsion and the contortion terms from the equation of the geodesic deviation (2.5) is obvious, since Riemannian geometry is a symmetric one ( we mean by a non-symmetric geometry, a geometry which admits a non-symmetric affine connection). Also, the disappearance of $ \tilde{Q}^{\alpha}_{.\mu\nu \sigma}$ part of the curvature term in this equation is due to the same reason.\\
2- The disappearance of the curvature term from equation (3.7) is due to the identical vanishing of the curvature (2.9), formed from the connection $\Gamma^{\alpha}_{. \mu \nu}$, as a consequence of the AP-condition. \\
3- It is clear from Table 2 that some quantum properties appear in the new set of path deviation equations. These properties appear in the parameters of different terms of the general equation (3.10), representing this set. These quantum properties appear without performing any known quantization scheme. \\
4- From the study of the path equations in AP-geometry (Table 1), the quantum properties appeared as a jumping coefficient of the torsion term. Now the study of the path deviation equations (Table 2) shows that, not only the coefficient of the torsion, but also the coefficient of the curvature, which jumps by a step of $\frac{1}{2}$.
The jumping of the coefficients of these terms are running in an inverse way. This result confirms a similar conclusion [7] obtained without using path deviation equations. \\
5- Although the jumping step, of the torsion term and the curvature term, is half integer; it clear from Table 2 that the jumping step of the contortion term is an integer. This may need further discussion.

It is worth of mention that the new set of path equations [5] has
been generalized [8], and applied successfully to discuss some
physical [9], [10] and astrophysical [11], [12] problems. These
applications may give the first evidence for the quantum structure
of space-time.
\section*{References}
{[1]} Wanas, M.I. and Bakry, M.A. (1995) Astrophys. Space Sci., {\bf{228}}, 239. \\
{[2]} Gurzadyan, V.G. (2003) {\it {Talk at XXII Solvey Conference on Physics}} (Delphi,

Nov.24-29,2001); astro-ph/0312523. \\
{[3]} Bazanski, S.L. (1989) J. Math. Phys., {\bf{30}}, 1018. \\
{[4]} Wanas, M.I. (2001) Stud. Cercet. \c Stiin\c t. Ser. Mat. Univ. Bac\u au {\bf{10}},297;

gr-qc/0209050. \\
{[5]} Wanas, M.I., Melek, M. and Kahil, M.E. (1995) Astrophys. Space Sci., {\bf{228}},

273; gr-qc/0207113. \\
{[6]} Wanas, M.I. and Kahil, M.E. (1999) Gen. Rel. Grav.,
{\bf{31}}, 1921;

 gr-qc/9912007. \\
{[7]} Wanas, M.I. (2003) Algebras, Groups and Geometries,
{\bf{20}}, 345a. \\
{[8]} Wanas, M.I. (1998) Astrophys. Space Sci., {\bf{258}}, 237; gr-qc/9904019. \\
{[9]} Wanas, M.I., Melek, M. and Kahil, M.E. (2000) Gravit.
Cosmol., {\bf{6}}, 319;

 gr-qc/9812085. \\
{[10]} Sousa, A.A. and Maluf, J.W. (2004) Gen. Rel. Grav., {\bf{36}}, 967; gr-qc/0301131. \\
{[11]} Wanas, M.I., Melek, M. and Kahil, M.E. (2002) Proc. MG IX,
p.1100, Eds.

V.G. Gurzadyan et al. (World Scientific Pub.);
gr-qc/0306086. \\
{[12]} Wanas, M.I. (2003) Gravit. Cosmol., {\bf{9}}, 109.

\end{document}